\RequirePackage{fix-cm}
\documentclass{svjour3} 
\smartqed 
\usepackage{graphicx} 
\usepackage{cite} 
\usepackage{amsmath,amssymb,amsfonts} 
\usepackage{algorithm,algorithmic} 
\usepackage{textcomp} 
\usepackage{xcolor} 
\usepackage{soul} 
\usepackage{balance} 
\usepackage{hyperref} 
\usepackage{orcidlink} 
\usepackage[mathlines,switch]{lineno} 
\usepackage{url} 
\usepackage{mathtools} 
\usepackage{multirow} 
\usepackage{pgfplots} 
\usepackage{tabularx} 
\usepackage{booktabs} 
\usepackage{caption} 
\usepackage{subcaption} 
\usepackage{footnote} 
\usepackage{float} 
\usepackage{adjustbox} 
\usepackage{colortbl}
\usepackage[listings,skins,breakable]{tcolorbox}

\def\BibTeX{{\rm B\kern-.05em{\sc i\kern-.025em b}\kern-.08em
    T\kern-.1667em\lower.7ex\hbox{E}\kern-.125emX}}

\makeatletter
\newcommand{\newlineauthors}{%
  \end{@IEEEauthorhalign}\hfill\mbox{}\par
  \mbox{}\hfill\begin{@IEEEauthorhalign}
}
\makeatother

\journalname{Journal}
\date{Received: date / Accepted: date}

\begin{document}

\title{Latent Space Class Dispersion: Effective Test Data Quality Assessment for DNNs}

\author{Vivek Vekariya\textsuperscript{1} \orcidlink{0009-0005-1482-7142} 
\and Mojdeh Golagha\textsuperscript{2} \orcidlink{123-456-7890} 
\and Andrea Stocco\textsuperscript{1,3} \orcidlink{0000-0001-8956-3894} 
\and Alexander Pretschner\textsuperscript{1,3} \orcidlink{0000-0002-5573-1201} 
}

\institute{
    \textsuperscript{1} TUM School of Computation, Information and Technology, Technical University of Munich, Munich, Germany. Email: vivek.vekariya@tum.de, andrea.stocco@tum.de, alexander.pretschner@tum.de\\
    \textsuperscript{2} Infineon Technologies AG, Munich, Germany. Email: mojdeh.golagha@infineon.com \\
    \textsuperscript{3} fortiss GmbH, Munich, Germany. Email: stocco@fortiss.org, pretschner@fortiss.org
}

\maketitle

\begin{abstract}
High-quality test datasets are crucial for assessing the reliability of Deep Neural Networks (DNNs). Mutation testing evaluates test dataset quality based on their ability to uncover injected faults in DNNs as measured by mutation score (MS). At the same time, its high computational cost motivates researchers to seek alternative test adequacy criteria. 
We propose Latent Space Class Dispersion (LSCD), a novel metric to quantify the quality of test datasets for DNNs. It measures the degree of dispersion within a test dataset as observed in the latent space of a DNN. 

Our empirical study shows that LSCD reveals and quantifies deficiencies in the test dataset of three popular benchmarks pertaining to image classification tasks using DNNs. Corner cases generated using automated fuzzing were found to help enhance fault detection and improve the overall quality of the original test sets calculated by MS and LSCD.

Our experiments revealed a high positive correlation (0.87) between LSCD and MS, significantly higher than the one achieved by the well-studied Distance-based Surprise Coverage (0.25). 
These results were obtained from 129 mutants generated through pre-training mutation operators, with statistical significance and a high validity of corner cases. 

These observations suggest that LSCD can serve as a cost-effective alternative to expensive mutation testing, eliminating the need to generate mutant models while offering comparably valuable insights into test dataset quality for DNNs.


\end{abstract}
\section{Introduction}

Deep Neural Networks (DNNs) are widely used for safety-relevant tasks, such as Automated Driving (AD)~\cite{nvidia-dave2}, medical diagnosis~\cite{zhang2018medical}, disease prediction~\cite{Zhao-nature}, or aircraft collision avoidance~\cite{Julian}. 
DNNs are trained to process complex input data (e.g., images) from large datasets, and their deployment in the real world depends on their generalization capabilities to correctly process unseen data. To this end, engineers use evaluation metrics such as accuracy, precision, recall, or mean squared error on test datasets unknown at training time~\cite{ml_testing_survey,2020-Riccio-EMSE}. 
Despite researchers' efforts to ensure the reliability and dependability of DNNs~\cite{testing_ml_industry_icse21, structured_verification_ml_rahul, ml_testing_survey,2020-Riccio-EMSE}, \emph{the quality of test datasets} remains a crucial factor in assessing their performance before deployment in real-world settings. This is because test datasets represent only a subset of the targeted operational domain, and they may fail to capture relevant properties of in-field inputs~\cite{2022-Stocco-TSE,guerriero2023iterative,DBLP:journals/corr/abs-2102-04287}.

Ideally, high-quality test datasets contain samples similar to the training set distribution and samples that represent rare situations occurring in the real world. Ensuring the adequacy and quality of test suites for DNNs is a crucial endeavor in software engineering~\cite{testing_ml_industry_icse21, structured_verification_ml_rahul, ml_testing_survey,2020-Riccio-EMSE}. 
In literature, researchers have addressed this problem through two main approaches: proposing adequacy criteria for DNNs and developing mutation testing techniques. Regarding adequacy criteria, various forms of neuron coverage have been introduced~\cite{neural_coverage_new,deepxplore}, though they are generally considered ineffective~\cite{nc_not_meaningful}. To improve upon this, surprise-based coverage criteria have been proposed~\cite{surprise_adquacy, surprise_adequacy_journal, refinement_sa, revisit_coverage}, which assess the level of ``surprise'' in test data by measuring its deviation from training set inputs.

Concerning the latter, Mutation Score (MS)~\cite{deep_mutation, deep_mutation_plus, deep_crime, mutation_operator_ese} quantifies the quality of test datasets by their ability to reveal various faults in DNNs. To calculate MS, pre-training mutation operators (MOs) that replicate actual DNN faults~\cite{taxonomy_faults} are used to generate mutant models~\cite{deep_mutation,deep_crime} as post-training mutation operators generally do not reflect real DNN faults~\cite{deep_crime}.
A test dataset with a higher MS is considered to be a good representative of the targeted operational domain, effectively identifying various faults and assessing the generalizability of DNNs. 
Both approaches, while popular, have shortcomings, particularly in terms of computational efficiency, which limits their scalability in practical applications. Specifically, Distance-based Surprise Coverage (DSC) is computationally intensive, with its computational time complexity growing with the size of the training dataset. Similarly, pre-training MOs require re-training of the original DNNs, making them computationally expensive and sensitive to the used hyperparameters~\cite{deep_crime,mutation_operator_ese}. As a result, researchers are actively investigating more computationally efficient test adequacy criteria as alternatives to DSC and MS.

To address these gaps, in this paper we propose \emph{Latent Space Class Dispersion (LSCD)}, a novel evaluation metric to evaluate test dataset quality in DNNs. Given a test dataset, LSCD quantifies class dispersion by analyzing the latent space representations of a trained DNN. It measures the mean distance of data points from their respective class centers, where each class center represents the mean position of training samples for a given class in the latent space.
We focus on the latent space of a DNN as the placement of latent space vectors dictates the DNN's behavior for a given input. Consequently, test adequacy criteria based on the latent space should provide an accurate measure of the DNN's generalizability.

Using LSCD, developers can quantitatively assess the test dataset quality. 
If the test data exhibits LSCD values similar to those of training data, this indicates that the DNN clearly identifies features of test data, referring to a high degree of similarity in the characteristics of the two datasets. While this resemblance reflects DNN's learning from the training data, it could hinder its generalizability. 
In this vein, LSCD can be used to characterize untested regions near the decision boundaries. This helps assess the DNNs' generalizability when encountering real-world samples within the operational design domain. Additionally, class-wise LSCD scores provide developers with valuable insights for debugging failure causes. 

We conducted an empirical study to evaluate LSCD using state-of-the-art DNNs on three benchmark datasets for image classification, MNIST~\cite{mnist}, SVHN~\cite{svhn}, and GTSRB~\cite{gtsrb}. First, we evaluate the effectiveness of the DNNs on their original test datasets using accuracy, distance-based surprise coverage ~\cite{surprise_adquacy,surprise_adequacy_journal,refinement_sa}, and LSCD. Then, we automatically generate corner cases using coverage-guided fuzzing for which DNNs achieve low accuracy but high DSC and LSCD. This suggests that latent space quantification, obtained using only the DNN under test, is more effective at characterizing low-quality datasets than traditional accuracy metrics. Also, we postulate that a high-quality test dataset, as evidenced by comprehensive coverage of the latent space, will also achieve a high mutation score across diverse mutated models. 
Therefore, we also studied the correlation between accuracy, DSC, LSCD, and mutation score on the original test and corner case datasets to validate and investigate the efficacy of LSCD. We found a strong positive correlation (0.87) between LSCD and MS derived from pre-training MOs. At the same time, DSC exhibits a weak positive correlation (0.24) with MS. A $\rho$-value of less than 0.03 for this correlation study rejects the null hypothesis, indicating that the correlation is not due to random behavior. We also found that the LSCD's computational time was significantly, on average, 36 times faster than DSC across all DNNs under test. This suggests that, in contrast to DSC, LSCD may be a computationally efficient alternative to MS, potentially reducing the need to create expensive mutant models for assessing test dataset quality.

In summary, this paper makes the following contributions:
\begin{itemize}
\item A novel evaluation metric called Latent Space Class Dispersion (LSCD) to quantify test data quality using the latent space of a trained DNN.
\item We offer a more comprehensive evaluation beyond the common practice of using adversarial attacks by leveraging fuzzing to generate real-world valid corner case data for evaluating test adequacy metrics and mutation scores.   
\item An empirical study to calculate the correlation between DSC, LSCD, and MS calculated on 129 mutant models belonging to 15 unique faults in three DNN image classifiers. 
\item Our findings suggest that LSCD can be used as an inexpensive surrogate measure for mutation scores for test dataset quality assessment. All results and data are available in our replication package~\cite{replication-package}.
\end{itemize}
\section{Background} \label{sec:background}

We first describe the latent space of a DNN and discuss the importance of maximizing its coverage for testing purposes. Then, we discuss an overview of distance-based surprise coverage ~\cite{surprise_adquacy,surprise_adequacy_journal,refinement_sa}, which pivots around maximizing latent space coverage and mutation scores, to discuss their importance in assessing the quality of test datasets.

\subsection{What is the Latent Space of a DNN?}

During training, the DNN learns complex feature representations from the input data and encodes them into a \textit{latent space}, i.e., a low-dimensional representation embedded within the layers of a DNN~\cite{GoodBengCour16}. Since each layer of the network transforms the input data into progressively more abstract and high-level representations, the final hidden layer before the output layer captures the most essential features of the input data, which are used for predictions~\cite{GoodBengCour16}.

The output layers for classification DNNs are usually softmax layers. In our work, we consider the last hidden layers (also referred to as logit layers) of a DNN as the latent space where the learning of the DNN is summarized for decision-making. 
Given an input, the latent space embeds all information that the DNN uses for inference. In other words, it encapsulates the maximum possible behaviors of the DNN. Therefore, a profound understanding of the latent space and its properties is essential for DNN testing and debugging.

\subsection{Why Leverage Latent Space to Enhance DNN Test Adequacy?} \label{sec:why_latent_space}

The DNN learns non-linear decision boundaries to separate input data distributions based on their characteristics. Ideally, a high-performing DNN has class regions that are well-separated from one another by non-linear classification/decision boundaries in the latent space. Consequently, we distinguish between latent space vectors located:

\begin{enumerate}
\item \textit{Near training data distribution:} When a latent space vector corresponding to a given test input is located far away from the decision boundary, near the latent space vectors corresponding to training data, the DNN clearly identifies features from the input, and it correctly classifies the input with high confidence~\cite{abs-2009-05835}. 
\item \textit{In the proximity of the decision boundary:} These latent space vectors can be easily manipulated by adding some noise (e.g., adversarial attacks~\cite{HuangPGDA17}, or image corruptions~\cite{hendrycks2019benchmarking,2025-Lambertenghi-ICST}) to cause incorrect classification. 
\item \textit{In sparse regions away from training distribution:} When a latent space vector is located in sparsely populated regions, away from training latent space vectors, the DNN fails to clearly identify the features from the input for reliable prediction. This could be due to an input with complex features not represented by similar examples during training, but it could still occur in the real world. 
\end{enumerate}

Test datasets mapped to the sparse region and near the decision boundaries have a higher potential to assess the generalizability of a DNN. We refer to these regions outside the original training data distribution in the latent space, which is not tested with the currently available test dataset as the \textit{untested regions of the latent space}. We aim to find an evaluation metric for DNNs that is informed by these untested regions of the latent space and seek to find more test data from this region. By doing so, we provide test engineers with a metric reflective of DNN's intrinsic learning capability while quantifying the test dataset quality. 

\subsection{Distance-based Surprise Coverage} \label{sec:dsa}

Kim et al.~\cite{surprise_adquacy} introduced Surprise Coverage (SC) criteria, which are designed to capture the \textit{``surprise''} behavior of test data that deviates from its training data, as observed by a DNN. Test data with complex features may be surprising as a DNN struggles to identify features for classification, making them valuable for testing. 
In this paper, we examine and compare our approach with distance-based SA, which is specifically designed for classification problems with DNNs---the main focus of our study. Below, we provide a brief overview of the key characteristics of distance-based SA.

Distance-based surprise coverage (DSC) uses latent space vectors to quantify test dataset quality. DSC considers a test input surprising if its latent space vector is located away from the training distribution---specifically, near the decision boundary (as discussed in (\autoref{sec:why_latent_space})). High-quality test datasets with complex features will result in higher values of DSC for reliable assessment of DNN's performance for real-world applications. In (\autoref{sec:dsc_calculation}), using an example, we will discuss the limitations of using DSC and how LSCD can mitigate these limitations. While Kim et al.~\cite{surprise_adquacy} used the term ``Activation Trace (AT)'', it is synonym to the more widely used term in the literature ``Latent Space Vector''. Thus, we will discuss ATs as latent space vectors in this paper.

\subsection{Mutation Score} 

A mutation score characterizes the quality of test datasets based on their ability to capture changes in mutant DNNs. The hypothesis is that high-quality test data should maintain their effectiveness under various fault conditions. To this end, various mutation operators broadly categorized as pre-training and post-training MOs, are used to generate mutant DNNs. The mutation score measures discrepancies between the predictions of a mutant model and the original model on the test dataset. A higher MS indicates good quality test data as it reflects changes in the DL system under test. However, to compute an MS, a test engineer must first generate mutant models. Due to their ineffectiveness in reflecting real faults in DNN, the post-training MOs are not considered practical ~\cite{deep_crime, mutation_operator_ese}. On the other hand, pre-training MOs are computationally expensive as they require re-training of the DNN. Despite their computational cost, pre-training MOs are essential in mutation testing due to a strong link to the real faults of DNNs ~\cite{deep_crime, mutation_operator_ese}. Thus, we study pre-training MOs derived from real faults of DNN in this paper and follow the mutation testing methodology from existing works~\cite{deep_mutation, deep_mutation_plus, deep_crime} for computing mutation scores.

Latent space-based test adequacy metrics such as DSC and LSCD reflect the learning capabilities of DNNs that are captured in their latent space. We hypothesize that such metrics can quantify the quality of test data that helps evaluate DNNs' generalizability and fault detection abilities. Thus, if a test dataset exhibits a higher DSC and LSCD, it should also correlate with higher mutation scores. To the best of our knowledge, the correlation of these test adequacy metrics, especially DSC with MS, has not been explored so far. Therefore, in this paper, we aim to bridge this gap by validating the efficacy of latent space-based metrics as a computationally efficient alternative to MS for assessing test dataset quality. 

\section{Methodology}\label{sec:approach}

In this section, we first introduce the latent space class dispersion metric and distance-based surprise coverage. Then, we describe a methodology to perform test dataset quality assessment using DSC and LSCD. Lastly, we describe the improvement in test dataset quality using Coverage-Guided (CGF) Fuzzing, as measured by LSCD and DSC. We also compare these metrics with the mutation score proposed for DL systems~\cite{deep_mutation, deep_mutation_plus, deep_crime}.

\subsection{Latent Space Class Dispersion: Definition and Interpretation}  \label{lsc_calculation}

Our work proposes an evaluation metric for DNNs that is informed by the untested regions of the latent space. To assess the topology of the latent space, we use the notion of dispersion from the class centers. The $\mathit{LSCD}_i$ is a per-class evaluation metric calculated using  \autoref{eq1}.

\begin{equation}\label{eq1}
	\begin{multlined}
		LSCD_i = \frac{{\sum\limits_{j=0}^{|s_i|} \text{dist}(\overrightarrow{c_i}, \overrightarrow{s_i}^{(j)})}}{{|s_i|}}\\
	\end{multlined}
\end{equation}

First, the centroid $\overrightarrow{c_i}$ of each ground-truth class is computed using the \emph{training set}. Assume a set of training or testing inputs $s_i$ for ground-truth class $i$ where the $j^{th}$ element maps to the latent space vector $\vec{s}_i^{(j)}$. We define $\mathit{LSCD}_i$ w.r.t. class $i$ on this data $s_i$ as the mean Euclidean distance of all corresponding latent space vectors from the centroid $\overrightarrow{c_i}$.  $\mathit{LSCD}$ calculated using \autoref{eq2} aggregates the $\mathit{LSCD}_i$ values for all $N$ classes. We discuss LSCD in the paper, as an improvement in all classes is indeed an improvement in overall dataset quality. 

\begin{equation}\label{eq2}
	\begin{multlined}
		LSCD= \frac{\sum\limits_{i=0}^{N} \mathit{LSCD}_i}{N}
	\end{multlined}
\end{equation}

Our proposed metric is a continuous-valued metric that does not rely on empirically calculated hyperparameters. In our empirical study (\autoref{sec:empirical-study}), we show that $\mathit{LSCD}$ can be used by test engineers to evaluate the quality of various test suites. The class-wise $\mathit{LSCD}_i$ scores are assessed for all output classes of the DNN under test, thereby reflecting the quality of feature identification for a given class. Thus, $\mathit{LSCD}_i$ scores can also be used by developers to define re-training strategies or robustness measures for DNN to avoid biased performance towards certain output classes.

\textbf{LSCD Interpretation:}
We discuss how test engineers can interpret $\mathit{LSCD}$ scores in conjunction with traditional evaluation metrics, e.g., accuracy. 
For example, high accuracy scores (e.g., $>90\%$) and low $\mathit{LSCD}$ scores (i.e., less than or close to the $\mathit{LSCD}$ scores of the training dataset) indicate that the majority of the test data points for all classes are \textit{in close proximity} of the class regions learned by the DNN. Analogous to DSC, the test data is \textit{not surprising} to the DNN. For a well-trained DNN, the assumption is justified that  every instance of the test data point has a low distance to its class center in this case (all latent space vectors belonging to the same class lie in close proximity to each other). On the other hand, in terms of test dataset quality, an LSCD value comparable to the training dataset indicates that (1)~the test data distribution is in line with the training data distribution but that the DNN may have overfitted to such data only, and (2)~the other regions of the latent space of the DNN are untested with the current test dataset. Thus, the test data exhibit poor diversity, threatening a reliable assessment in the real world because the in-field data are unlikely to always resemble the training data.

In this paper, we also scrutinize corner case data generated using the CGF framework (\autoref{sec:cfg}) that shows improved test dataset quality through increased LSCD. Corner case data in image classification refers to the set of images that the DNN under test is likely to misclassify. A high $\mathit{LSCD}$ value of corner case data compared to the original train and test set indicates that most data points are located relatively far away from the class regions learned by the DNN, and therefore, when using the original test dataset in the \textit{previously untested region of the latent space}. The combination of low accuracy scores (e.g., $<5\%$) and higher $\mathit{LSCD}$ scores when comparing train and test sets suggests that the corner case dataset with a higher LSCD score has sufficient diversity, as the DNN has failed to identify the specific features from them enough to generalize to rare or highly diverse test data. 
 
\subsection{Comparison of LSCD with Distance-based Surprise Coverage} \label{sec:dsc_calculation}
Unlike LSCD, DSC is calculated in multiple steps ~\cite{surprise_adquacy}. Firstly, a new test input $x$ is mapped to the nearest neighbor $x_a$ in its output class $c_x$ using \autoref{dsa1}. Then, the Euclidean distance between the latent space vector of $x$ and its nearest found neighbor $x_a$ is calculated using \autoref{dsa2}. Here, $\overrightarrow{\alpha_N}(.)$ refers to the latent space vectors for a given input. 

\begin{equation}\label{dsa1}
x_a = \arg\min_{D(x_i) = c_x} \|  \overrightarrow{\alpha_N}(x) - \overrightarrow{\alpha_N}(x_i) \|
\end{equation}

\begin{equation}\label{dsa2}
\text{dist}_a = \| \overrightarrow{\alpha_N}(x) - \overrightarrow{\alpha_N}(x_a) \|
\end{equation}

Second, the nearest neighbour $x_b$ for $x$ is found from other output classes $C \setminus \{c_x\}$ using \autoref{dsa3}. Using \autoref{dsa4}, the Euclidean distance between the latent space vector of $x_a$ and $x_b$ is calculated. Here, $C$ is the total number of output classes. Finally, DSC is computed as a ratio of distances calculated from the output class to the nearest input in another class using \autoref{dsa5}. 
\begin{equation}\label{dsa3}
x_b = \arg\min_{D(x_i) \in C \setminus \{c_x\}} \| \overrightarrow{\alpha_N}(x_a) - \overrightarrow{\alpha_N}(x_i) \|
\end{equation}

\begin{equation}\label{dsa4}
\text{dist}_b = \| \overrightarrow{\alpha_N}(x_a) - \overrightarrow{\alpha_N}(x_b) \|
\end{equation}

\begin{equation}\label{dsa5}
\text{DSC($x$)} = \frac{\text{dist}_a}{\text{dist}_b}
\end{equation}

The DSC for all inputs in test dataset $X$ is defined as \autoref{dsa6}.
DSC segments are divided into  \( k \) segments using an upper bound $U$ and buckets \( B = \{b_1, b_2, \ldots, b_k\} \) that divide the interval \( (0, U] \).

\begin{equation}\label{dsa6}
\text{DSC($X$)} = \frac{\left| \left\{ b_i \mid \exists x \in X : DSC(x) \in \left( U \cdot \frac{i-1}{k}, U \cdot \frac{i}{k} \right] \right\} \right|}{k}
\end{equation}

\begin{figure}[t]
 \centering
 \includegraphics[trim=0.8cm 0.8cm 0.8cm 0.8cm, clip=true, width=0.35\linewidth, keepaspectratio]{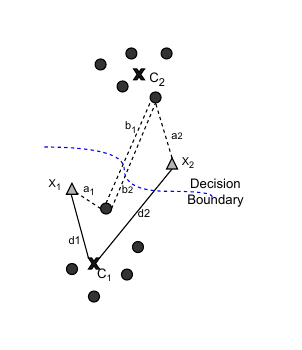}
 \caption{Illustration of DSC and LSCD calculation. Black dots represent latent space vectors from the training set for classes 1 and 2 with their centers C$_1$ and C$_2$. New test inputs X$_1$ (correctly classified) and X$_2$ (falsely classified), both belonging to class 1, are shown to demonstrate the metric calculations.}
  \label{fig:dsc_lscd}
\end{figure}

\autoref{fig:dsc_lscd} illustrates the fundamental differences in computing DSC and LSCD for two new test inputs X$_1$ and X$_2$ belonging to the same class. However, the DNN under test classifies X$_1$ correctly and X$_2 $ incorrectly. First, DSC identifies the nearest latent space vectors from the training dataset and calculates the Euclidean distance a$_1$ and a$_2$ for X$_1$ and X$_2$, respectively. In the second step, DSC calculates inter-class distances b$_1$ and b$_2$ between these test inputs and the nearest vectors from other classes. As a result of this, DSC calculated using \autoref{dsa5} results in $\frac{a_1}{b_1} \approx \frac{a_2}{b_2}$. Furthermore, the presence of outliers in the training dataset can potentially skew the results as the first step in DSC calculation relies on nearest-neighbor distances. This factors leads to a key limitation: DSC fails to distinguish between correctly classified samples and misclassified samples. Another limitation of using DSC is its computational complexity, which increases with the size of the training data. 

On the other hand, LSCD takes a centroid-based simple approach, offering potential efficiency advantages. LSCD calculates the Euclidean distance between each test input's latent representation and the centroid of its ground truth class. In \autoref{fig:dsc_lscd}, d$_1$ and d$_2$ represent the distance of X$_1$ and X$_2$ to their ground-truth class center. In a well-trained DNN, d$_1$ (for the correctly classified X$_1$) is strictly smaller than d$_2$ (for the misclassified X$_2$). In this way, LSCD can effectively differentiate between correctly and incorrectly classified samples. Moreover, LSCD inherently mitigates the impact of outliers using centroids. Since centroids are the mean representations, they are less influenced by the outliers. We empirically verify this by evaluating the quality of corner cases from fuzzing using DSC and LSCD and analyzing its correlation with MS and accuracy in \autoref{sec:empirical-study}. 

\subsection{Correlation Analysis of DSC and LSCD with Mutation Score}\label{sec:ms}
\begin{figure}[H]
  \centering
  \includegraphics[width=0.9\linewidth, keepaspectratio]{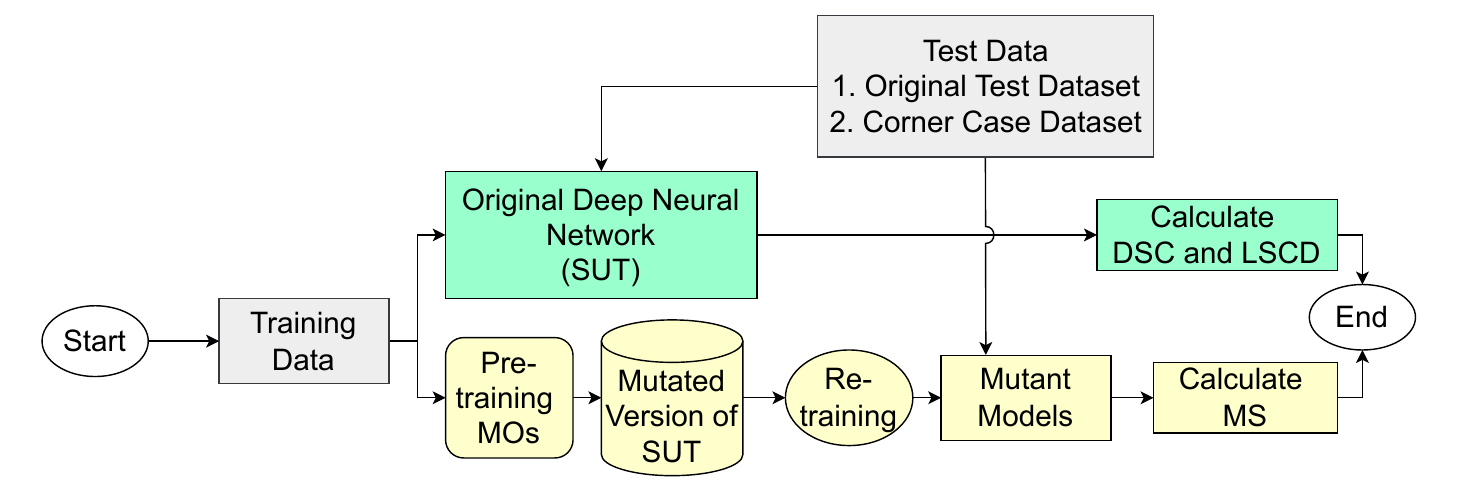}
  \caption{Our framework to evaluate test dataset quality based on Distance-based Surprise Coverage and Latent Space Class Dispersion (pipeline highlighted in green blocks). We compare our approach with the Mutation Score (highlighted in the pipeline with yellow colored blocks).} 
  \label{fig:our_modified_framework}
\end{figure}

Following the insights presented so far, a high-quality test dataset, as indicated by its latent space-based evaluation score (here, DSC and LSCD), has a higher potential to verify the generalization capabilities of DNNs. Similar to these metrics, a high MS also indicates a good quality test dataset that helps assess the generalization and fault detection capabilities of DNNs. Thus, we compute and study the correlation of DSC and LSCD with mutation scores using mutant DNNs derived from pre-training MOs. \autoref{fig:our_modified_framework} overviews the individual steps, which include test dataset quality evaluation using three different methods, namely using DSC, LSCD, and mutation score.

Using our proposed method, we first calculate standard evaluation metrics such as accuracy and test dataset quality using DSC and LSCD. High accuracy alone might create a false notion of trustworthiness in DNN's performance. This is because the latent space metrics (relative to the training data) suggest that the DNN effectively identifies features in the current test dataset similar to those in the training set, indicating a lack of diversity in test data. This may lead to a DNN that might not generalize well to real-world datasets with complex features. To verify our hypothesis that these factors are crucial to assessing dataset quality, we evaluate the dataset quality of corner cases identified using fuzzing. If gains in DSC/LSCD and drops in accuracy are observed, i.e., the generated corner cases have higher DSC/LSCD than the original test dataset, it indicates that DNN finds them \textit{surprising} and struggles to identify features in them. These corner case data can enhance the quality of the original test dataset as they can be mapped to the sparse and near decision boundary regions in the latent space, making them more suitable for verifying the generalization of DNNs.

To evaluate the effectiveness of DSC and LSCD under various fault conditions and select the most suitable criteria, we also compute their correlation with MS on the mutant DNNs obtained from pre-training MOs. A strong positive correlation would mean that the test engineers can use a metric with the highest correlation as a computationally efficient alternative to mutation score for evaluating test data quality, thereby eliminating the need to generate mutant DNNs.We also measured the computation time required to compute DSC and LSCD to provide insights into their computational efficiency. Post-training MOs tend to be unrealistic as they directly change the trained model, which is unlikely to happen in real-world settings~\cite{deep_crime,teasma}. Also, they lack precise mapping to actual DNN faults ~\cite{deep_crime,taxonomy_faults}. Thus, we focus on pre-training MOs that mimic real faults in DNNs.

\subsection{Corner Case Generation using Coverage-Guided Fuzzing}\label{sec:cfg}

We use Coverage-Guided Fuzzing (CGF) to sample corner case data to improve the original test dataset quality, which was evaluated using DSC, LSCD, and mutation scores. The test dataset that maximizes the latent space coverage can be achieved through two primary approaches: (1)~upsampling latent space vectors from the untested latent space region and generating an input using a decoder or (2)~using some automated tool for generating inputs with desired properties that can increase the latent space coverage.

Existing work by Mani et al.~\cite{metrics_ibm} focuses on the first case, generating corner case images by upsampling and traversing through the latent space vectors rather than randomly selecting. However, the validity of the generated images strictly depends on the quality of the decoder used to reconstruct the image from the latent space vector, which may fail to reliably reconstruct the image from the latent space vector. Moreover, the use of a decoder often results in the generation of many invalid images that only contain noise and are semantically different from the original test data~\cite{metrics_ibm}. Indeed, latent space exploration for test generation is still an open research topic~\cite{2023-Riccio-ICSE}.

Therefore, in our work, we focus on the second case and remove the dependency on a decoder-based DNN to generate corner case images. To this end, we use CGF, a promising technique for the automated generation of corner cases of DNNs using various coverage criteria like Neuron Coverage (NC), k-Multisection Neuron Coverage (k-MNC), Neuron Boundary Coverage (NBC)~\cite{deephunter, deepxplore, deepgauge}. 

\begin{figure}[t]
  \centering
  \includegraphics[width=0.90\linewidth, keepaspectratio]{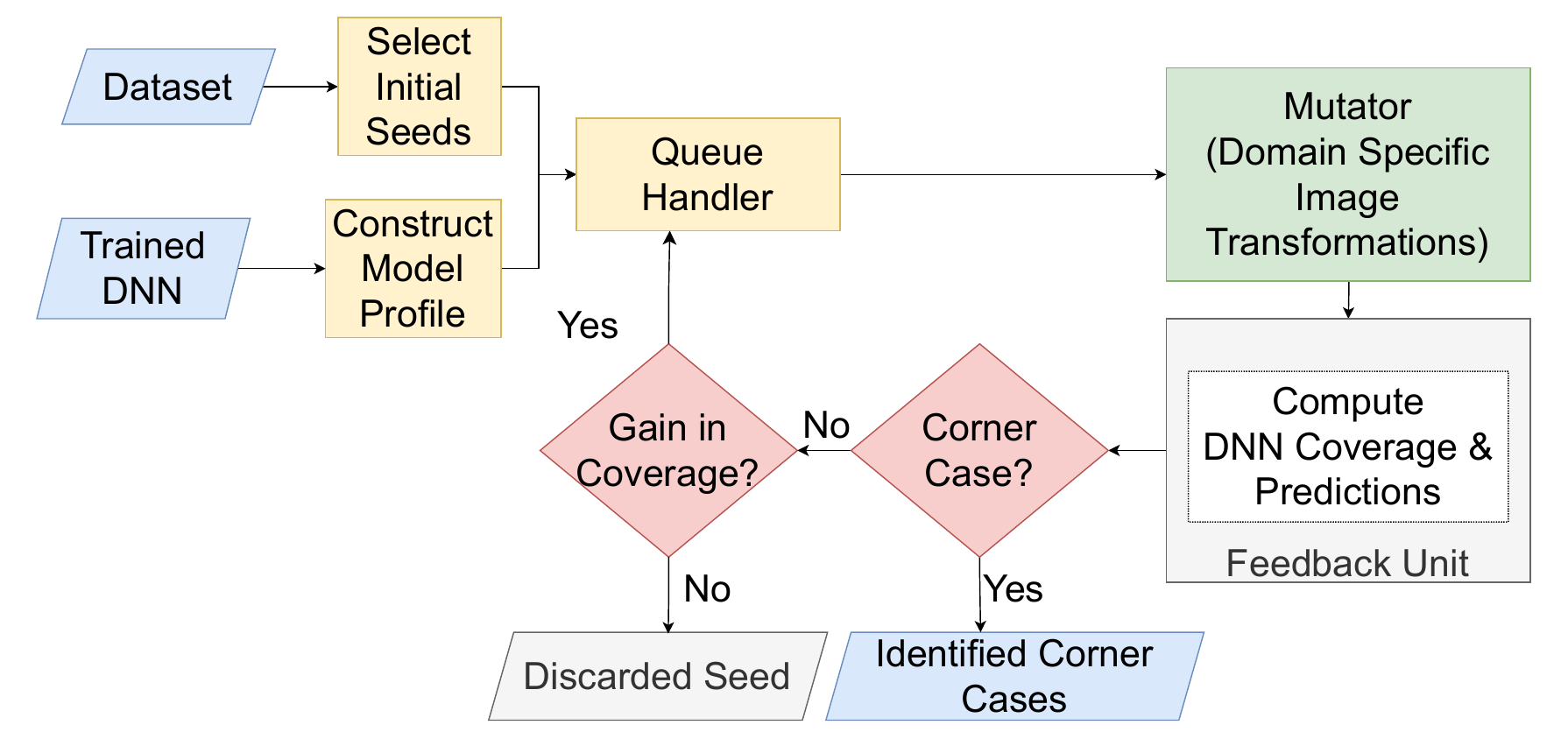}
  \caption{Coverage-Guided Fuzzing}
  \label{fig:cgf}
\end{figure}

\begin{table}[t]
\renewcommand{\arraystretch}{1.5}
\Large
\captionsetup{justification=justified}
\caption{Summary of hyper-parameters for coverage-guided fuzzing used to sample corner case data}
\label{tbl:fuzzing_parameters}
\begin{adjustbox}{width=\columnwidth}
\begin{tabular}{l l l}
\toprule
\textbf{Fuzzing step} & \textbf{Requirement} & \textbf{Hyper-parameters}\\
\midrule
Selection of Initial Seeds & To maximize fuzzing efficiency & Correctly classified images in classification \\ 
\midrule
Coverage Criteria & To maximize DNN's behavior & NC, k-MNC, NBC with hyper-parameters 0.75, 1500, and 10 respectively\\ 
\midrule
Mutator & To add realistic transformation & Widely used domain-specific transformation~\cite{hendrycks_framework,2022-Stocco-TSE} \\ 
\midrule
Corner Case Identification & Criteria to select corner cases & Falsely classified images\\ 
\midrule
Number of iterations & Stopping criteria for fuzzing & 15,000 iterations for all classification DNNs\\ 
\bottomrule
\end{tabular}
\end{adjustbox}
\end{table}

The use of coverage-guided fuzzing helps generate diverse test data by leveraging annotated test data and identifying hidden defects and weaknesses of the DNN under test. The coverage criteria used in CGF aim to verify the maximum possible behaviors of DNN against various input conditions, which can be useful in increasing the diversity of the original test data. Our experimental study empirically validates this hypothesis (\autoref{sec:empirical-study}). 

The CGF framework is depicted in \autoref{fig:cgf}.
The fuzzing process starts with the selection of initial seeds on which the DNN has high performance as reflected by evaluation metrics. Then, the mutator applies a transformation to the selected seed image. If a transformed image helps increase the currently used coverage criterion, it is added to the queue as it helps explore the maximum possible behavior of the DNN. Corner cases are identified if the DNN makes false predictions on these transformed images. A detailed summary of the hyper-parameters associated with fuzzing in our case study can be found in \autoref{tbl:fuzzing_parameters}. We run fuzzing until the coverage criterion used reaches a saturation level because continuing with the same criteria might lead to diminishing returns in fuzzing efficiency. This is because the same criteria are likely to identify fewer and fewer corner cases over time. We calculate and compare the LSCD and standard evaluation metrics on fuzzing data to demonstrate the enhancement in test dataset quality as indicated by these metrics. 

\section{Empirical Study} \label{sec:empirical-study}

This section outlines our empirical study designed to evaluate the efficacy of our proposed metric. To this aim, we aim to answer the following research questions:

\noindent
\textbf{RQ\textsubscript{1} (Correlation Study):} 
\textit{What is the correlation between our proposed metric LSCD, DSC, and MS in test data quality assessment?}

We performed an empirical study to evaluate which metric among LSCD and DSC can be used as a computationally efficient substitute for mutation scores by computing their correlation with MS using Pearson correlation coefficient~\cite{pearson_wiki}. To do so, we first quantify the dataset quality of both the original test data and corner case data using DSC, LSCD, and MS obtained from a set of mutant models. We then calculate the Pearson correlation coefficient along with the statistical significance test.

\noindent
\textbf{RQ\textsubscript{2} (Computational Time Required):} 
\textit{What is the relative improvement in computation time for our metric LSCD compared to the original and refined DSC approaches?}

Various studies~\cite{refinement_sa, surprise_adequacy_journal} highlight computational limitations when using the original single-threaded DSC implementation~\cite{surprise_adquacy} in practice. Although Weiss et al. in~\cite{refinement_sa} propose a refinement of DSC to make it computationally faster in multi-threading settings, LSCD by design is computationally faster. It requires distance calculation only to the class center in a single step. Nonetheless, we also investigated the computational advantages of using LSCD by comparing its computational time with DSC (original and refined) under both single- and multi-threaded setups.

\noindent
\textbf{RQ\textsubscript{3} (Corner Case Validity):}
\textit{What is the validity of the corner case data as assessed by automated input validators?}

As higher DSC, LSCD and MS values can be achieved with noisy and ``unrealistic'' data, it is essential to assess whether increases in LSCD and MS are associated with the generation of inputs that are valid in the sense of being ``realistic'', that is, recognizable by humans in the input domain \cite{2023-Riccio-ICSE}. To retain the realism and reliability of our correlation study, we hence assess whether the corner case data are valid according to an automated input validator from the literature.

\subsection{Datasets and Deep Neural Networks} \label{sec:dnn_architecture}

Our empirical study includes datasets of increasing diversity and complexity, along with various DNN architectures, to increase the reliability of our findings. Particularly, we consider the popular MNIST~\cite{mnist}, SVHN~\cite{svhn}, and GTSRB datasets~\cite{gtsrb} for image classification with DNNs.
\autoref{tbl:dataset} reports statistics about the datasets, such as the number of output classes and their size (divided into training, validation, test set, and identified corner case images using CGF).

\begin{table}[H]
\caption{Overview of Datasets}
\label{tbl:dataset}
\begin{tabular}{ccccccccc}
\toprule
&\textbf{No. of} & \multicolumn{6}{c}{\textbf{Number of Images}} \\
\cmidrule{3-8}
\textbf{Dataset} & \textbf{Output}  & \textbf{Training}& \textbf{Validation}& \textbf{Test} & \multicolumn{3}{c}{\textbf{Corner Case Images using}}\\
\cmidrule{6-8}
 & \textbf{Classes}  & & & & NC & k-MNC & NBC\\
\midrule
MNIST & 10 & 48,000 & 12,000 & 10,000 & 14,129 & 12,868 & 14,006\\
SVHN & 10 & 62,268 &10,989 & 26,032  & 14,390 & 14,216 & 14,339\\
GTSRB & 43 & 35,610 & 3,870 & 12,631 & 13,327 & 14,476 & 14,249\\
\bottomrule
\end{tabular}
\end{table}

MNIST~\cite{mnist} is a dataset of hand-written digits depicted as gray-scale images, often used in DNN testing works. SVHN~\cite{svhn} is a large-scale dataset from Google Street View images that comprises real-world images of house numbers. SVHN is helpful in studying advanced classification DNNs due to its complexity and diversity. GTSRB is a benchmark dataset for German traffic sign recognition. 
It is a feature-rich dataset consisting of images with various illumination changes, rotations, and occlusions likely to be encountered in actual driving. Thus, it serves as a benchmark to test traffic sign classification DNNs. 

 We use state-of-the-art image classification networks based on LeNet-5~\cite{lenet_5} and Spatial Transformer Networks (STNs) available from previous work~\cite{gtsrb_repo} for our study. The choice of LeNet-5 for MNIST helps understand the latent space and its properties under less complex settings. On the other hand, using the same STN architecture for SVHN and GTSRB allows us to understand the latent space for datasets of varying complexity and output dimensions. These DNNs are trained using the Adam~\cite{adam} optimizer with a learning rate scheduler and early stopping criteria for 50 epochs (MNIST) and 100 epochs (SVHN and GTSRB). The architectures of the DNNs and their performance using test data can be found in \autoref{tbl:dnn_architecture} and \autoref{tbl:dnn_performance_1}, respectively.

\begin{table}[t]
\renewcommand{\arraystretch}{1.5}
\Large
\centering
\caption{Details about DNN architectures used for the empirical study}
\label{tbl:dnn_architecture}
\begin{adjustbox}{width=\columnwidth}
\begin{tabular}{lll}
\toprule
\textbf{Network} & \textbf{Layer} & \textbf{Parameters/Description} \\
\midrule
\multirow{4}{*}{LeNet-5} & Convolutional 1 & 6 filters, 5x5 kernel, stride 1, ReLU, Average Pooling \\
& Convolutional 2 & 16 filters, 5x5 kernel, stride 1, ReLU, Average Pooling \\
& Fully Connected 1 & Flatten and 120 neurons, ReLU \\
& Fully Connected 2 & 84 neurons, ReLU \\
& Output (Logit Layer + Softmax) & 10 neurons (MNIST), Softmax \\
\midrule
\multirow{8}{*}{Spatial Transformer Network} & Localization Network & Convolutional layers (learn affine transformation parameters) \\
& Grid Generator & Creates sampling grid \\
& Sampler & Samples input feature map \\
& Convolutional 1 & 100 filters, 5x5 kernel, LeakyReLU, BatchNormalization \\
& Convolutional 2 & 150 filters, 3x3 kernel, LeakyReLU, BatchNormalization\\
& Convolutional 3 & 250 filters, 3x3 kernel, LeakyReLU, BatchNormalization\\
& Fully Connected 1 & Flatten and 350 neurons, ReLU \\
& Output (Logit Layer + Softmax) & 10 and 43 neurons (SVHN and GTSRB respectively), Softmax \\
\bottomrule
\end{tabular}
\end{adjustbox}
\end{table}

\subsection{Latent Space Vector Selection}\label{sec:latent-vector-selection}

To calculate LSCD and DSC, first, we need to select latent space vectors. A fully connected layer is a common type of output layer used in DNNs for the image classification task. The size of the fully connected layer is equal to the number of output classes. Thus, the latent space vector is a vector of a constant size corresponding to the given input image. We used the latent space vector derived from the last fully connected layer (often referred to as the logit layer as shown in \autoref{tbl:dnn_architecture}) for LSCD and DSC calculation. 

\subsection{Procedure and Metrics}

\subsubsection{RQ\textsubscript{1}: Correlation Study}\label{sec:rq1-increase}

To study the correlation of LSCD and DSC with mutation score, we start with the baseline evaluation of the DNN under test. This involves calculating accuracy, DSC, and LSCD using the original train and test datasets. Then, we generate the corner case dataset using NC, k-MNC, and NBC as coverage criteria during fuzzing. We also compute accuracy, LSCD, and DSC on the generated corner cases. 
The DSC results are calculated using \autoref{dsa6} using bucket size $k$=1000, which is commonly used in the literature~\cite{surprise_adequacy_journal, refinement_sa}. 
The LSCD is calculated using \autoref{eq2}.

To investigate which metric between DSC and LSCD is better suited for evaluating dataset quality and replacing mutation scores, we study their correlation with mutation scores, which assess test dataset quality based on their ability to capture faults in DNNs. We consider the fault categories proposed by Humbatova et al.~\cite{taxonomy_faults}. We identified 15 unique faults that belong to 3 out of 5 top-level categories of the final taxonomy. The two omitted categories pertain to the use of incorrect hardware and tensors used to train DNNs, which do not apply in our implementation. Moreover, we used selected faults to define MOs corresponding to them. We trained 129 mutant models for the classification DNNs using these MOs. A detailed description of fault category, MOs, and corresponding hyper-parameters can be found in \autoref{tbl:mutation_operator_table}.

\begin{table}[!t]
\renewcommand{\arraystretch}{1.2}
\captionsetup{justification=justified}
\caption{Fault names, mutation operators, and their associated hyper-parameters employed to generate mutant models~\cite{taxonomy_faults}.}
\label{tbl:mutation_operator_table}
\begin{adjustbox}{width=\textwidth}
\begin{tabular}{l l l l l} 
\toprule
\textbf{Sr. No.} & \textbf{Fault Group} & \textbf{Fault Name} &\textbf{Mutation Operator} & \textbf{Hyper-Parameters Used}  \\
\toprule
1       & \multirow{5}{*}{Training - Hyper parameters} & \multirow{2}{*}{Suboptimal batchsize}  & Increase Batch Size  & 64    \\
2       &  &    & Decrease Batch Size          & 2048, 4096 \\
\cmidrule{3-5}
3       &  & \multirow{2}{*}{Suboptimal learning rate}   & Decrease Learning   Rate     & 0.0001 \\
4       &   &    & Increase Learning   Rate     & 0.01, 1  \\
\cmidrule{3-5}
5       &   & Wrong optimizer   function    & Change Optimizer& Adagrad   \\
\midrule
6       & \multirow{2}{*}{API}   & \multirow{2}{*}{Missing API Call} & Remove Zero Grad & API call     \\
7       &    &   & Remove Call & API call\\
\midrule
8       & \multirow{4}{*}{Training - Training Data}    & \multirow{2}{*}{Low quality training data}   & Add Training Noise           & 0.25, 0.75, 0.9  \\
9       &  &     & Remove Samples               & 0.25, 0.75, 0.9 \\
\cmidrule{3-5}
10      &   & Overlapping output   classes     & Make Classes Overlap         & 0.25, 0.75, 0.9     \\
11      & & Wrong labels for   training data & Change Most Labels (in \%) & 25, 60  \\
\midrule
\multirow{2}{*} {12}      & Model - Activation Function& Wrong activation / & Change Activation & Tanh, LogSigmoid,   Sigmoid\\      
 & & Missing Relu activation function &   \\
\midrule
13      & \multirow{2}{*}{Model - Layer Properties}   & Wrong filter size for convolutional layer   & Layer Size & Output channels changed\\
14      &  & Bias needed in a  layer    & Remove Bias                  &  according to architecture \\
\midrule
\multirow{2}{*} {15}   & Model - Missing/Redundant     & Missing dropout layer   & Dropout   & 0.25, 0.8    \\ 
 & /Wrong Layer & & &  \\
\bottomrule
\end{tabular}
\end{adjustbox}
\end{table}

We calculate MS using \autoref{eq3}, a modified version of the formula proposed in DeepMutation++~\cite{deep_mutation_plus}.

\begin{equation}\label{eq3}
	\begin{multlined}
		MS_{m'} = \frac{{|\left\{ t \mid t \in T \wedge m(t) \neq m'(t) \right\}|}} {|T|}
	\end{multlined}
\end{equation}

Here, MS measures the discrepancy between the predictions of a mutant model $m'$ and the original model $m$ on the tests $t\in T$ (as defined in Equation~\ref{eq3}). The only modification compared to DeepMutation++ is not aggregating the MS for all mutant models. In this way, we calculate individual accuracy, DSC, LSCD, and MS values for all mutant models.

Following the calculation of these metrics, we compute the correlation between accuracy, DSC, LSCD, and MS along with a statistical significance test using the Pearson correlation coefficient~\cite{scipy, pearson_wiki}. A strong positive correlation between MS and latent space-based metrics (LSCD and DSC) would support the proposition that these metrics offer a computationally inexpensive way to quantify test dataset quality, as they do not rely on mutant models. The metric with the strongest correlation coefficient would then be considered the more suitable alternative among LSCD and DSC. We used the classical $p$-value of $0.05$ is as a threshold to indicate whether the correlation is statistically significant.

\subsubsection{RQ\textsubscript{2}: Computational Time Required}\label{sec:rq2-computational-time}

The original DSC calculation uses single-thread execution, which makes it computationally expensive~\cite{surprise_adquacy, refinement_sa}. Thus, Weiss et al.~\cite{refinement_sa} proposed a refined multi-threaded DSC calculation, which is computationally faster. To explore LSCD's behavior under different threading models, we compared the computational time required to calculate LSCD with both the single- and multi-threaded DSC implementations. As both metrics require latent space vectors, we pre-calculated and stored all latent space vectors to ensure a fair comparison focused solely on metric calculation time.     

\subsubsection{RQ\textsubscript{3}: Corner Case Validity}\label{sec:rq3-validty}
The results of this correlation study can be adversely affected by the inclusion of invalid and inherently poor-quality corner case images. A higher MS and LSCD value can also be achieved when using noisy images that are not recognizable to humans as well. Thus, we also assess and report the validity and quality of the identified corner cases using an automated input validator, SelfOracle~\cite{2020-Stocco-ICSE}. It is a distribution-aware input validator for image data, which has shown high agreement with the human assessment of validity in a large comparative study about test input generators for deep learning~\cite{2023-Riccio-ICSE}. 

SelfOracle is based on a Variational Autoencoder (VAE)~\cite{An2015VariationalAB} trained to minimize the reconstruction error between the original and reconstructed inputs. After training, the VAE will have a higher reconstruction error on invalid data (highly corrupted data that fall out-of-distribution), whereas it will exhibit lower reconstruction errors for valid, in-distribution data~\cite{2023-Riccio-ICSE}. SelfOracle leverages probability distribution fitting to find the threshold that discriminates valid inputs from invalid ones~\cite{2023-Riccio-ICSE}. It fits a Gamma distribution to the training data through maximum likelihood estimation~\cite{mle-gamma}. Probability distribution fitting provides a statistical model for the reconstruction errors of nominal data. This statistical model can be used by the tester to configure the threshold with the desired rate of false alarms (i.e., inputs that belong to the nominal dataset but are classified as invalid). 

Using the same hyperparameters of the original study~\cite{2023-Riccio-ICSE}, we trained SelfOracle using the training set of each dataset for 100 epochs, with a learning rate of 0.001. The network uses the Adam optimizer to minimize the mean squared error (MSE) loss. The latent space dimension was 1024. 
After training, we applied SelfOracle to reconstruct all corner case images for all datasets using the false alarms rate of the original study ($0.01$\%). Inputs with a reconstruction error above this threshold are considered invalid, while inputs below it are considered valid~\cite{2023-Riccio-ICSE}.
\section{Results} \label{Results}

\subsection{RQ\textsubscript{1} (Correlation Study)}

\autoref{tbl:dnn_performance_1} reports the performance of DNNs for image classification across all three datasets. The accuracy on the original test dataset is $98.60$\%, $92.83$\%, and $99.71$\% for MNIST, SVHN, and GTSRB datasets, respectively. In contrast, performance significantly declined when evaluated using the sampled corner case data, as expected. The mean accuracy on GTSRB corner cases dropped to the lowest $1.48$\%, while the accuracy on MNIST and SVHN corner cases decreased to $3.35$\% and $9.56$\%, respectively. The dataset quality quantification of the original test and sampled corner case data using LSCD and DSC explains the reason for these varying performances. 

\begin{table}
\renewcommand{\arraystretch}{1.2}
\Large
\captionsetup{justification=justified}
\caption{Summary of performance evaluation of DNNs using Accuracy and test dataset quality assessment using DSC and LSCD. The values in the bracket for corner case data are in percentage points, which indicates the drop/gain in the evaluation metric compared to the original test data}
\label{tbl:dnn_performance_1}
\begin{adjustbox}{width=\columnwidth}
\begin{tabular}{c c c c c} 

\toprule

\multirow{2}{*}{\textbf{Dataset}} & \multirow{2}{*}{\textbf{Data Split}} & \multicolumn{3}{c}{\textbf{Evaluation Metrics}}\\ 

\cmidrule{3-5} 

& & \textbf{Accuracy (\%)} & \textbf{DSC} & \textbf{LSCD}\\ 

\midrule

\multirow{4}{*}{\textbf{MNIST}}  & Test data & 98.60 & 0.43 & 25.27 \\
& Corner Case (NC) & 3.93 \textbf{(-94.64)} & 0.61  \textbf{(+0.18)} & 57.17 \textbf{(+31.90)}\\  
& Corner Case (k-MNC) & 3.04 \textbf{(-95.56)} & 0.57  \textbf{(+0.14)} & 57.79 \textbf{(+32.52)}\\  
& Corner Case (NBC) & 3.08 \textbf{(-95.52)} & 0.59 \textbf{(+0.16)} & 57.56 \textbf{(+32.29)}\\  
\midrule
\multirow{4}{*}{\textbf{SVHN}}  & Test data & 92.83 & 0.62 & 6.14 \\
& Corner Case (NC) & 9.46 \textbf{(-83.37)} & 0.53  \textbf{(-0.09)} & 10.73 \textbf{(+4.59)}\\  
& Corner Case (k-MNC) & 9.46 \textbf{(-83.37)} & 0.51  \textbf{(-0.102)} & 10.74 \textbf{(+4.60)}\\  
& Corner Case (NBC) & 9.76 \textbf{(-83.07)} & 0.49 \textbf{(-0.122)} & 10.73 \textbf{(+4.59)}\\ 
\midrule
\multirow{4}{*}{\textbf{GTSRB}}  & Test data & 99.71 & 0.58 & 55.71 \\
& Corner Case (NC) & 1.61 \textbf{(-98.10)} & 0.77  \textbf{(+0.22)} & 186.73 \textbf{(+131.02)}\\  
& Corner Case (k-MNC) & 1.50 \textbf{(-98.21)} & 0.80  \textbf{(+0.22)} & 188.31 \textbf{(+132.60)}\\  
& Corner Case (NBC) & 1.33 \textbf{(-98.38)} & 0.79 \textbf{(+0.22)} & 189.09 \textbf{(+133.38)}\\  
\bottomrule

\end{tabular}
\end{adjustbox}
\end{table}

The DSC values for the MNIST train, test, and corner case (NC) datasets are 0.001, 0.43, and 0.73, respectively. A higher DSC value indicates that the corresponding inputs are more surprising to the DNN. This is because inputs with complex features tend to have latent space vectors far from the training distribution placed in the sparse or near decision boundary region. Therefore, DSC values for the misclassified images in corner case data are the highest. The test data, which achieves high accuracy, also shows a moderately high DSC of 0.51 compared to the training dataset due to the inherent nature of DSC. This is likely due to test data, while correctly classified, existing at the boundaries of their training class distributions. This vicinity combined with inter-class distances computed using \autoref{dsa4}, results in a higher DSC compared to simple, well-centered training examples. The consistent DSC-based test dataset quality evaluation in SVHN and GTSRB datasets demonstrates its failure to differentiate between the original test data and corner case data, which can have relatively high DSC values. Surprisingly, the corner case data for SVHN resulted in a mean DSC value of 0.53, which is less than the original test data of 0.62. This could be due to the presence of outliers in training distribution, which can skew the DSC calculations. 

Using LSCD, we evaluate the topology of the latent space with respect to class centers, which represent the mean of the training distribution. The LSCD values of MNIST train, test, and corner case data (NC) are 25.10, 25.27, and 57.17. The similar LSCD values of train and test data suggest that they exhibit similar class dispersion around the derived class centers. In contrast, the LSCD value of corner case data is significantly higher than the original train and test data, which demarcates the dispersion of misclassified data from the correctly classified one. The LSCD values for SVHN train, test, and corner case data (NC) are 6.08, 6.14, and 10.73 respectively. This contrasts the dataset quality evaluation using DSC, where corner case data were less surprising than the original test data. This implies that the use of centroids in LSCD calculation effectively reduces the impact of training dataset outliers, unlike DSC. Datasets with higher LSCD are good quality test data having a higher potential to verify a DNN's generalizability as they assess sparse regions of latent space that lie strictly outside the training distributions. 
Furthermore, in contrast to DSC, LSCD provides a computationally simpler assessment of the latent space without the calculation of inter-class distances.

\begin{table}
\captionsetup{justification=justified}
\caption{Answer to RQ1: Correlation analysis using Pearson Correlation Coefficient}
\label{tbl:correlation_analysis}
\centering
\begin{adjustbox}{width=\columnwidth, center}
\begin{tabular}{cccccc}
\toprule
\textbf{Metric Comparison} & \textbf{Dataset} & \textbf{No. of Mutant DNNs} & \textbf{Correlation} & \textbf{p-value} \\
\midrule
DSC vs. MS & MNIST & 40 & \cellcolor{gray!20} 0.574 & 0.00 \\
& SVHN & 42 & \cellcolor{red!20} 0.020 & 0.00 \\
& GTSRB & 47 & \cellcolor{red!20} 0.159 & 0.00 \\
\cmidrule{2-5}
& \textbf{Average} & - & \cellcolor{red!20} \textbf{0.251} & \textbf{0.00} \\
\midrule
LSCD vs. MS & MNIST & 40 & \cellcolor{green!20} 0.771 & 0.03 \\
& SVHN & 42 & \cellcolor{green!20} 0.878 & 0.03 \\
& GTSRB & 47 & \cellcolor{green!20} 0.975 & 0.03 \\
\cmidrule{2-5}
& \textbf{Average} & - & \cellcolor{green!20} \textbf{0.874} & \textbf{0.03} \\
\midrule
Accuracy vs. MS & MNIST & 40 & -0.942 & 0.03 \\
& SVHN & 42 & -0.929 & 0.03 \\
& GTSRB & 47 & -0.992 & 0.03 \\
\cmidrule{2-5}
& \textbf{Average} & - & \cellcolor{green!20} \textbf{-0.954} & \textbf{0.03} \\
\midrule
Accuracy vs. LSCD & MNIST & 40 & -0.786 & 0.03 \\
& SVHN & 42 & -0.825 & 0.03 \\
& GTSRB & 47 & -0.978 & 0.03 \\
\cmidrule{2-5}
& \textbf{Average} & - & \cellcolor{green!20} \textbf{-0.863} & \textbf{0.03} \\
\midrule
Accuracy vs. DSC & MNIST & 40 & -0.540 & 0.00 \\
& SVHN & 42 & -0.123 & 0.00 \\
& GTSRB & 47 & -0.135 & 0.00 \\
\cmidrule{2-5}
& \textbf{Average} & - & \cellcolor{red!20} \textbf{-0.266} & \textbf{0.00} \\
\bottomrule
\end{tabular}
\end{adjustbox}
\end{table}

\autoref{tbl:correlation_analysis} presents the correlation analysis, including the result of statistical tests, for accuracy, mutation score, DSC, and LSCD across the three datasets. We recognized that the DSC calculation is sensitive to the bucket size $k$ used in \autoref{dsa6}. To mitigate any potential bias in the correlation study, we performed DSC calculations across various bucket sizes in the range of [100, 1000] with a step size of 100. The correlation values reported for DSC in \autoref{tbl:correlation_analysis} is the mean Pearson correlation across various bucket sizes. 

\begin{figure}[t]
    \centering
    \begin{subfigure}{0.45\textwidth}
        \centering
        \includegraphics[width=\textwidth, keepaspectratio]{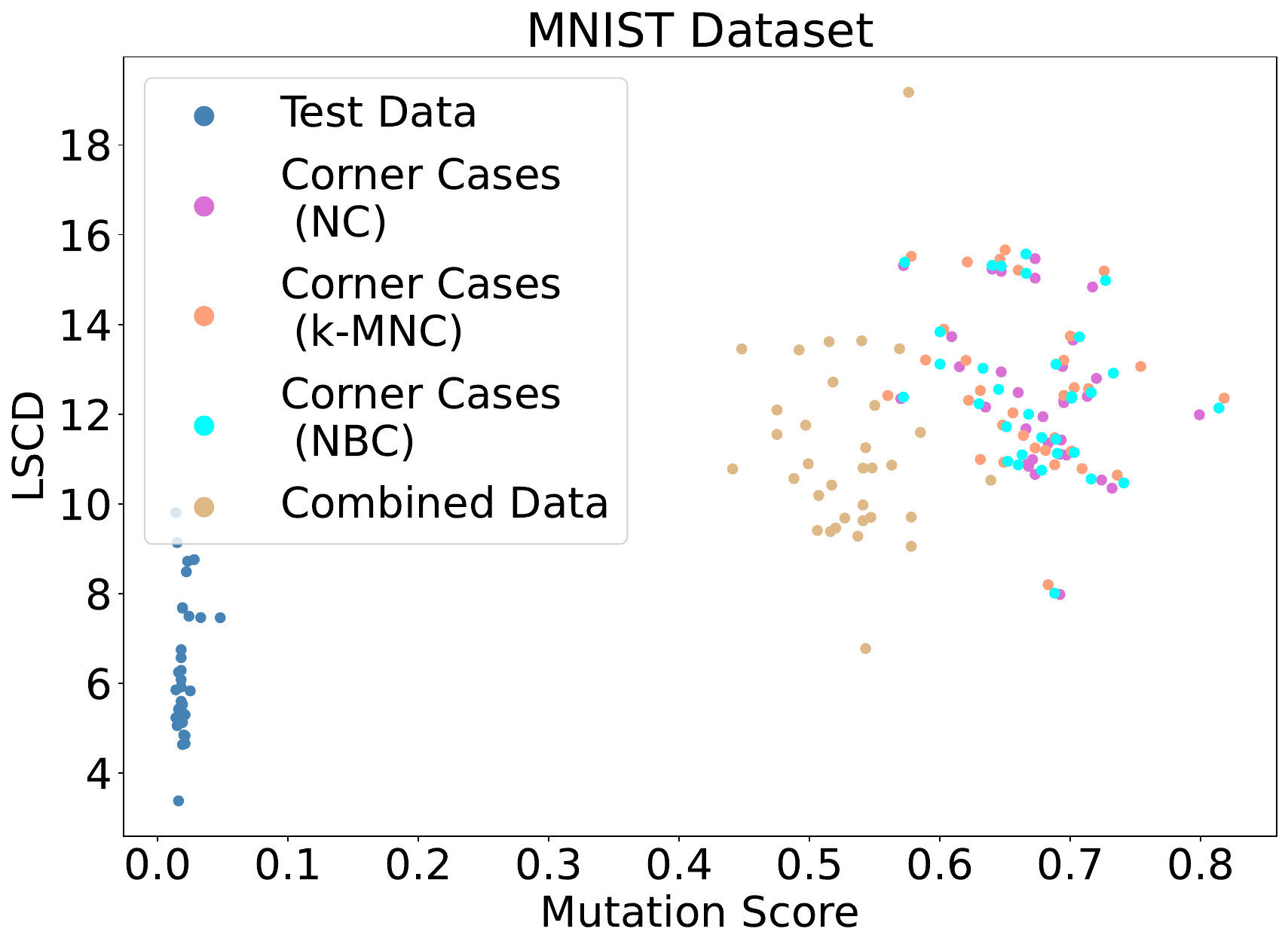}
        \caption{}
        \label{fig:fig1}
    \end{subfigure}
    \hfill
    \begin{subfigure}{0.45\textwidth}
        \centering
        \includegraphics[width=\textwidth, keepaspectratio]{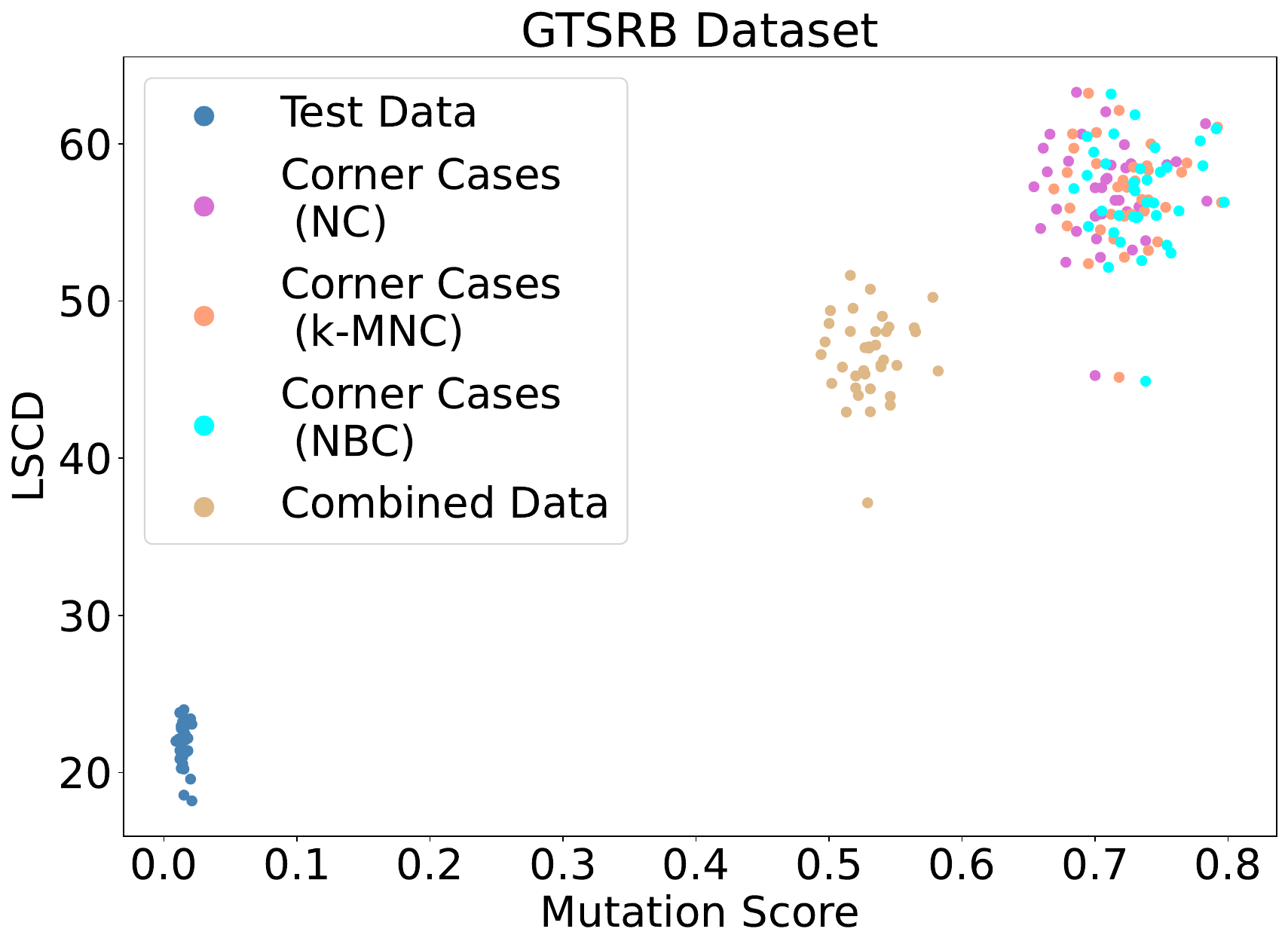}
        \caption{}
        \label{fig:fig3}
    \end{subfigure}

        \begin{subfigure}{0.45\textwidth}
        \centering
        \includegraphics[width=\textwidth, keepaspectratio]{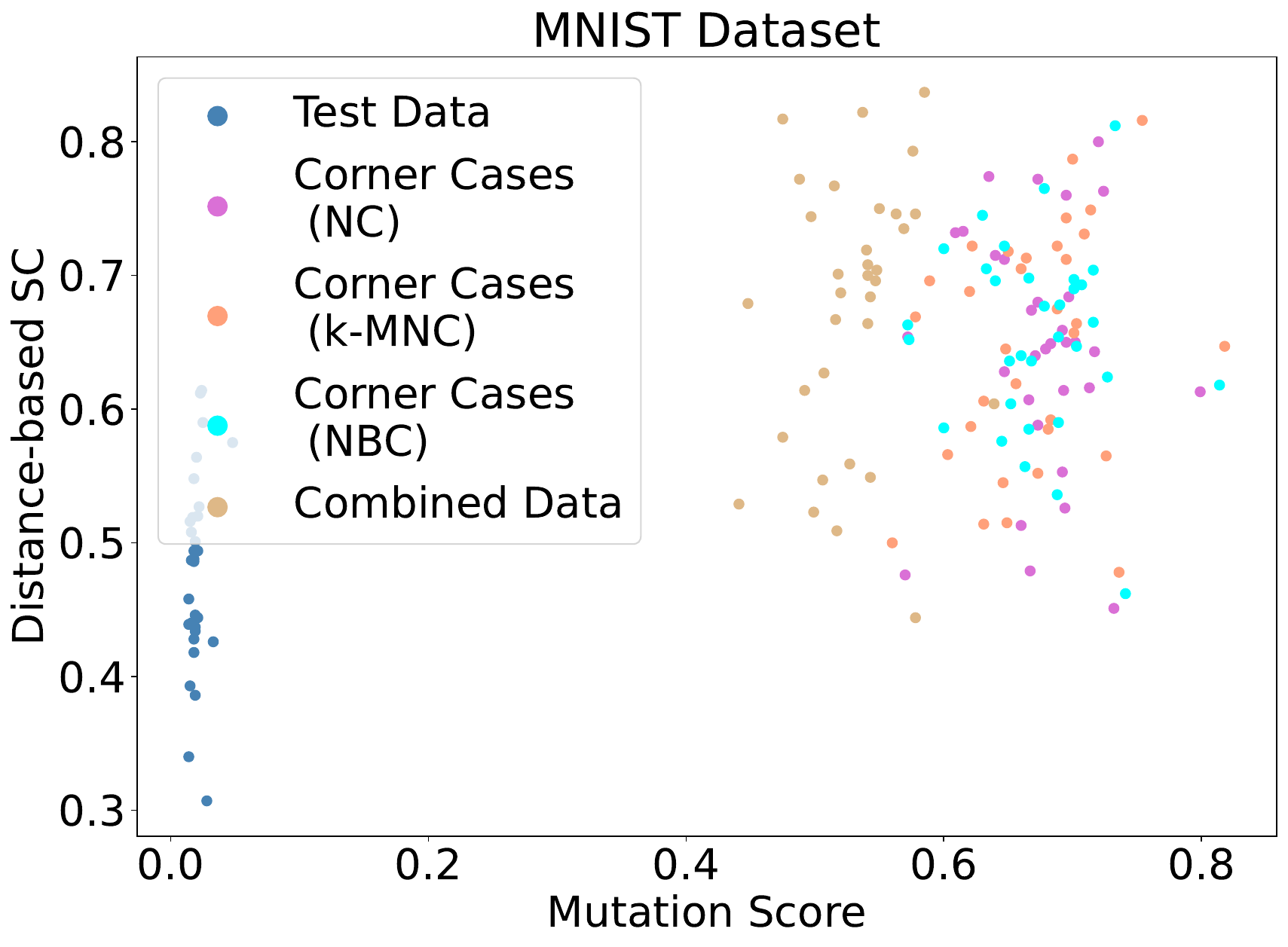}
        \caption{}
        \label{fig:fig4}
    \end{subfigure}
    \hfill
    \begin{subfigure}{0.45\textwidth}
        \centering
        \includegraphics[width=\textwidth, keepaspectratio]{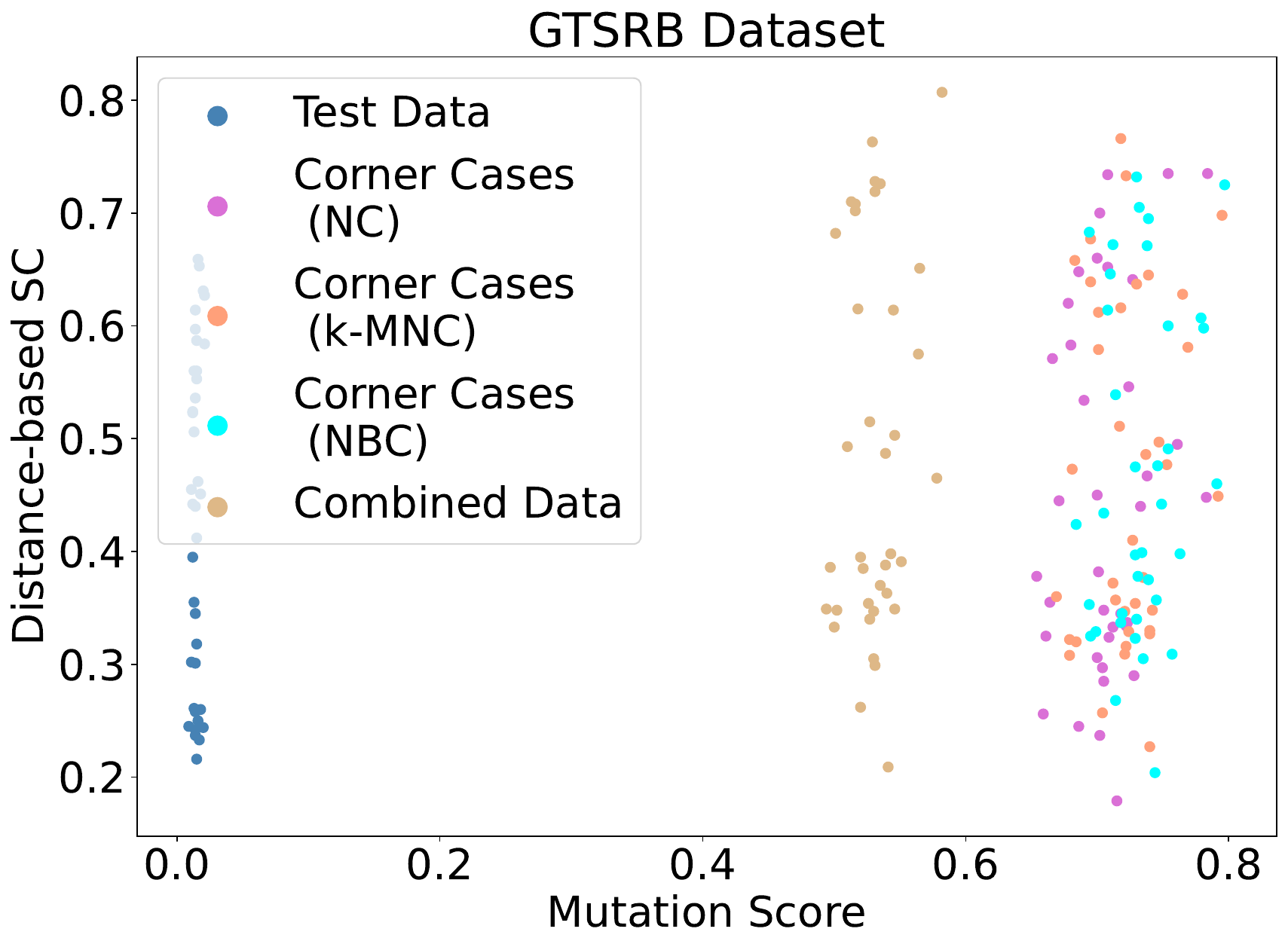}
        \caption{}
        \label{fig:fig6}
    \end{subfigure}
    
    \caption{LSCD (a-b) and DSC (b-c) v/s MS values across all mutant models.}
    \label{fig:lscd_ms_figs}
\end{figure}

As previously established, LSCD clearly distinguishes between test data that are correctly and incorrectly classified, a distinction DSC struggles to make. Consequently, the LSCD exhibits a significantly stronger inverse correlation with accuracy, yielding a Pearson correlation of -0.863, compared to DSC's weak negative correlation of -0.266 across all mutant models and datasets. Our main goal is to find the more suitable candidate to approximate the mutation score. To this aim, we observed a strong positive correlation between MS and LSCD (0.847), contrasting with the weak positive correlation between MS and DSC (0.251). \autoref{fig:lscd_ms_figs} shows how DSC, LSCD, and MS values vary across all mutant models and datasets. 

We would like to highlight that for simpler networks and datasets, such as LeNet-5 and MNIST, the correlation between DSC, LSCD, and mutation score is very close. Specifically, for MNIST, DSC demonstrates a moderately positive correlation with MS (0.574), while LSCD shows a strong positive correlation (0.771), indicating a less complex latent space. However, a significantly stronger relationship between LSCD and MS emerges with the use of more sophisticated networks like STN with SVHN and GTSRB datasets. The statistical significance of all reported correlation values is confirmed by p-values less than 0.03, allowing us to reject the null hypothesis and conclude that the results are not due to chance.

\tcbset{breakable = false, enhanced jigsaw, left=1mm, right=1mm, bottom=1mm, boxsep=1mm}
\begin{tcolorbox}[title=\textbf{Answer to RQ1:  Correlation Study}]
\textit{LSCD shows a significantly stronger positive correlation with MS (0.874) than DSC (0.251), due to DSC's inability to discern between true and falsely classified samples. 
Also, the correlation between LSCD and MS increases with the use of complex networks like STN and datasets like SVHN and GTSRB.
}
\end{tcolorbox}

\subsection{RQ\textsubscript{2} (Computational Time Required)}

We used an Ubuntu 22.04 machine with AMD Ryzen 2920X 12-Core Processor, NVIDIA GeForce RTX 2070 SUPER, and 128GB RAM in our experiments. We monitor the computational time required for LSCD and DSC calculations across all mutant models under single and multi-thread settings of both criteria.

The original DSC implementation~\cite{surprise_adquacy} employed single-threaded computation, which calculates image-wise DSC by calculating the distance against the entire training set using \autoref{dsa1} and \autoref{dsa2}. As shown in \autoref{tbl:compute_time}, single-threaded execution to compute DSC takes 3,491 minutes in total, across all datasets and mutant models. The multi-threaded implementation reduced the computation time for DSC across all mutant models and datasets to approximately 100 minutes and 30 seconds, achieving a 97.11\% speedup. This performance improvement is due to the parallelization of latent vector distance computations~\cite{refinement_sa}, which yielded the expected performance improvement. This also highlights the inevitable necessity of multi-threaded execution for efficient DSC computation.

Our proposed metric LSCD requires distance calculations to a single class center only. Thus, using single-threaded execution, vectorized implementation allows the user to compute LSCD across all mutant models and datasets in just 2 minutes and 45 secs. In contrast, multi-threading introduces an overhead due to the creation of multiple CPU threads for each distance calculation, resulting in a substantially longer computation time of 35 minutes and 5 seconds. This suggests that LSCD doesn't necessarily need multi-threading and is optimized for single-thread execution. However, when comparing the computational time required for LSCD and DSC, LSCD is significantly faster than DSC by 36.55 times.

\begin{table}[]
\captionsetup{justification=justified}
\caption{Answer to RQ2: Runtime comparison (minutes:seconds) for calculating DSC and LSCD in single-threading and multi-threading implementations.}
\label{tbl:compute_time}

\begin{adjustbox}{width=\columnwidth}

\begin{tabular}{c c c c c c} 
\toprule

\textbf{Dataset} & \textbf{No. of Mutants} & \multicolumn{2}{c}{\textbf{Single Threading}} & \multicolumn{1}{c}{\textbf{Multi Threading}}\\

\cmidrule(lr){3-5}  

& &  \textbf{LSCD} & \textbf{DSC} & \textbf{DSC} \\ 
\midrule
MNIST & 40 & \cellcolor{green!20} 00:27 & 889:00  & \cellcolor{red!20} 15:44 \\
SVHN & 42 & \cellcolor{green!20} 00:37 & 1,669:00  & \cellcolor{red!20} 29:38 \\
GTSRB & 47 & \cellcolor{green!20} 01:41 & 929:00  & \cellcolor{red!20} 55:08 \\
\midrule
Total & 129 & \cellcolor{green!20} 02:45 & 3,487:00  & \cellcolor{red!20} 100:30\\
\bottomrule
\end{tabular}
\end{adjustbox} 
\end{table}

\tcbset{breakable = false, enhanced jigsaw, left=1mm, right=1mm, bottom=1mm, boxsep=1mm}
\begin{tcolorbox}[title=\textbf{Answer to RQ2:  Computational Time Required}]
\textit{Single-threaded LSCD computation is, on average, 36.55 times faster than multi-threaded DSC computation across all mutant models and datasets.
}
\end{tcolorbox}

\subsection{RQ\textsubscript{3} (Corner Case Validity)}

The inclusion of poor-quality corner case data poses a potential threat to the reliability of our correlation study in RQ1, as noisy data can also result in a higher MS, DSC, and LSCD. We generated a substantial number of corner cases through fuzzing, such as 41,003 images for MNIST, and similar amounts for SVHN and GTSRB. We used these images to study the correlation between LSCD and DSC with MS. To ensure the reliability of our correlation study, we calculated the validity of corner case images. \autoref{tbl:vae_validitiy} illustrates the autoencoder-validity results for corner case images generated using fuzzing. The table reports the number of valid and invalid images along with the validity rate. Across all datasets, the average validity rate is 99.56\%, with very few corner cases reported as invalid, which built strong trust in the correlation study discussed in RQ1. 

\begin{table}[t]
\renewcommand{\arraystretch}{1.2}
\captionsetup{justification=justified}
\caption{Answer to RQ3: Corner Case validity assesment using SelfOracle~\cite{2020-Stocco-ICSE}.}
\label{tbl:vae_validitiy}

\begin{adjustbox}{width=\columnwidth}

\begin{tabular}{c c c c c c} 

\toprule

\multirow{2}{*}{\textbf{Dataset}} & \multirow{2}{*}{\textbf{Coverage  Criteria}} & \multicolumn{4}{c}{\textbf{Corner Case Data Evaluation Metrics}} \\

\cmidrule(lr){3-6}  

& & \textbf{Total No. of Images} & \textbf{No. of Valid Images} &  \textbf{No. of Invalid Images} & {\textbf{Validity (\%)}}\\ 

\midrule

\multirow{3}{*}{\textbf{MNIST}} & NC & 14,129	& 14,070	 & 59 & 99.58 \\
& k-MNC & 12,868	& 12,779 & 89	& 99.30 \\
& NBC & 14,006	& 13,946 &	60	& 99.57 \\
\midrule
\multirow{3}{*}{\textbf{SVHN}} & NC & 14,390 & 14,307	& 83 & 99.42 \\
& k-MNC  & 14,216 & 14,126 & 90 & 99.36 \\
& NBC & 14,339 &	14,256 & 83	& 99.42 \\
\midrule
\multirow{3}{*}{\textbf{GTSRB}} & NC & 13,327 & 13,305 & 22	& 99.83 \\
& k-MNC  & 14,476 & 14,451 & 25 & 99.82 \\
& NBC  & 14,249 & 14,216 & 33 & 99.76  \\

\midrule
Average & - & - & - & - & \textbf{99.56}\\

\bottomrule

\end{tabular}
\end{adjustbox} 
\end{table}

\tcbset{breakable = false, enhanced jigsaw, left=1mm, right=1mm, bottom=1mm, boxsep=1mm}
\begin{tcolorbox}[title=\textbf{Answer to RQ3:  Corner Case Validity}]
\textit{The identified corner case images used to study correlation exhibit a high validity rate for all datasets (99.56\% on average), according to a state-of-the-art automated input validator.
}
\end{tcolorbox}

\subsection{Discussion}

We identified three key explanations for the varying magnitude of LSCD values across datasets and the classification DNNs. First, the size of the latent space is a characteristic inherent to the DNN, which is shaped during training. Second, LSCD scores are impacted by the complexity of the architecture of the DNN under test. Spatial transformer networks used for SVHN and GTSRB are more complex networks, leading to a dense latent space compared to LeNet5 for MNIST. Last, the size of latent space is directly related to the number of output classes. Consequently, LSCD values are smaller for MNIST (max. 57.79) and SVHN (max. 10.74) with the fewer output classes and highest for GTSRB (max. 188.73) with the most. These properties are indeed helpful when focusing on the latent space-based characterization of DNN's behavior as they cater to the specificity of DNN under test.

Existing coverage criteria like NC, k-MNC, NBC calculate DNN structural coverage but are immaterial to dataset quality improvements achieved when used during fuzzing. Indeed, existing work by Harel-Canada et al.~\cite{nc_not_meaningful} suggests that achieving $100$\% NC is possible with only a few inputs while still being able to identify corner cases after achieving 100\% NC. Therefore, latent space-based quantification would help complement such techniques as it can better evaluate the test suite quality. In our experiments, the corner cases always led to an increase in LSCD and DSC. However, due to the computation of inter-class distance, DSC has significant computational overhead and limited ability to differentiate between correctly and incorrectly classified images. Therefore, LSCD can serve as a computationally efficient alternative. A test suite with higher LSCD effectively targets untested latent space regions. For the reliable acceptance of our newly proposed metric LSCD, we also studied the correlation of LSCD and DSC with MS using mutant models from pre-training MOs. 

The relevance of post-training MOs in the application domain is still an ongoing research topic~\cite{empirical_evaluation_mo, teasma} as they directly modify the internal structure of a trained DNN, which is not likely to happen in a real-world application. The use of latent space vectors for LSCD allows us to apply it to various DNNs and datasets by selecting latent space vectors according to the DNN under test.

\subsection{Threats to Validity}\label{sec:ttv}

\subsubsection{Internal validity} 
The increase in latent space coverage, as measured by LSCD and DSC, depends on the transformation techniques used during mutations in CGF. We have carefully chosen the domain-specific transformations that could occur in a real-world situation and are suitable for AD. The latent space coverage scores could change if other transformations are applied. 

\subsubsection{External validity}
We used a limited number of DNNs and datasets in our evaluation, which poses a threat in terms of the generalizability of our results. We tried to mitigate this threat by choosing popular datasets and existing state-of-the-art DNN models such as LeNet and Spatial Transformer Networks, which achieved competitive scores~\cite{lenet_5, gtsrb_repo}. 

\subsubsection{Reproducibility}
Concerning reproducibility, we make our code and artifacts available in our replication package~\cite{replication-package}. Specifically, it contains the mutation testing framework for classification DNNs, code for generating mutant models using pre-training MOs and calculating MS, DSC, and LSCD and the corner case data available with their associated metadata and validity assessment report.
\section{Related Work} \label{Related Work}
\subsection{Test Adequacy Metrics for DNNs}

Various test adequacy criteria and test input generators based on them have been proposed for DNNs~\cite{surprise_adquacy,deepxplore,deepgauge, deephunter, deeproad, deeptest}. Structural coverage criteria such as neuron coverage have been shown less useful for test data quality measurement~\cite{nc_not_meaningful}. On the other hand, latent space-based test adequacy metrics such as surprise coverage enjoy wider applicability. As studied by Weiss et al. in~\cite{refinement_sa}, surprise coverage has wide applicability and has already been cited in more than 105 scientific papers. Due to its computational complexity, refinement and computationally efficient versions of these metrics have been proposed~\cite{refinement_sa}. However, to the best of our knowledge, surprise coverage is only studied in the context of adversarial attack data, which is less meaningful considering their occurrence in the real world. Furthermore, to the best of our knowledge, we have found limited research exploring the correlation between surprise coverage metrics and mutation scores, which assess test dataset quality based on their ability to expose faults under varying conditions. 

Different from distance-based surprise coverage metrics, we propose LSCD, which is a more systematic way to define the latent space properties by considering the composition and nature of class regions centered around class centers in the latent space. Due to its simplicity, LSCD is computationally faster than DSC, the metric most closely related to our approach. We empirically evaluated and compared LSCD and DSC by examining their correlation with mutation scores. In this paper, we leverage corner case data generated through fuzzing, a technique widely employed by industry practitioners for finding corner case behaviors~\cite{ml_testing_survey, testing_ml_industry_icse21}. These corner case data have a high probability of occurrence in real-world scenarios. We evaluated our metric in the context of image classification using different DNN architectures.

\subsection{Mutation Testing for DNNs}

Mutation testing is well-studied for DL systems inspired by mutation testing in software systems~\cite{deep_mutation, deep_mutation_plus, deep_crime}. DeepMutation, DeepMutation++, and DeepCrime~\cite{deep_mutation, deep_mutation_plus, deep_crime} successfully proposed frameworks to create mutant models of trained DNNs and compute MS. Mutation score is used to rank various test suites based on their ability to reveal the faults in mutant models. DeepMutation++~\cite{deep_mutation_plus} uses post-training MOs, which lack realism to real-world applications as studied in~\cite{deep_crime, teasma}. To ensure that the findings in our study were relevant to real-world DNN behavior, we focused on realistic pre-training MOs. In DeepCrime~\cite{deep_crime}, the authors derived the pre-training MOs based on the real-world fault categories and computed MS. The choice of MOs to inject the faults and the availability of various formulas to compute MS make it more challenging to adapt it for practical uses. On top of that, the computationally expensive re-training steps involved in generating mutant model raise a serious question about their usefulness. Thus, we propose and explore the most suited alternative to MS. We introduced and evaluated LSCD as a promising replacement for MS. In our study, we found that LSCD outperforms DSC in terms of computational efficiency and has a stronger positive correlation to MS. The taxonomy of real faults in DL systems~\cite{taxonomy_faults} describes 92 unique faults arising from various sources leading to the faulty behaviors of DNNs. We chose 15 unique faults and created mutant models using corresponding pre-training MOs for our correlation study.

\subsection{Test Data Generation for DNNs}

The existence of a labeled and robust test oracle is very limiting. This has fostered the development of many automated test data generation techniques for DNN like DeepHunter~\cite{deephunter}, DeepXplore~\cite{deepxplore}, DeepGauge~\cite{deepgauge}. These methods are used in industry~\cite{ml_testing_survey, testing_ml_industry_icse21} to test the robustness properties of DNN models. This means identifying corner cases in which DNN exhibits poor performance. 
In our work, we use the CGF technique similar to DeepHunter to sample the corner cases for test data quality enhancement. 
While DeepHunter transforms the input image by using metamorphic relation strategies based on affine and pixel value transformations~\cite{deephunter}, in our paper, we use domain-specific transformations such as weather transformations (e.g., fog, rain), image transformations (e.g., saturate, contrast), blurring effects (e.g., defocus blur, gaussian blur)~\cite{hendrycks_framework} which are more realistic for the image classification task. Also, we reported the validity of sampled corner cases assessed by the SOTA automated validator which is missing in DeepHunter~\cite{deephunter}.

\section{Conclusion and Future Work} \label{Conclusion_future_work}

The test dataset quality plays a pivotal role in trusting the performance of DNN reflected in evaluation metrics. Mutation score measures test dataset quality based on their fault-revealing nature. Despite its promising insights, the practical application of mutation scores is limited due to the significant computational overhead required to generate mutant models. In this paper, we evaluated latent space-based test adequacy metrics as potential candidates to replace MS. We proposed a novel evaluation metric called LSCD to quantify test data quality considering the latent space of DNNs. We study the correlation of MS, LSCD, and DSC, a closely related and well-studied latent space metric. 

In our study considering two DNN classifiers and three datasets, we found that LSCD is strongly positively correlated with mutation scores, reflected in the Pearson correlation coefficient of 0.87. In contrast, DSC has a weaker positive correlation of 0.25 with MS. 
LSCD's computational time across 129 mutant models and all datasets was 36.55 times faster than DSC. This computational advantage and a stronger positive correlation highlight LSCD as a better and computationally efficient alternative than DSC to mutation scores. 

In future work, we aim to evaluate the effectiveness of LSCD and compare it with conventional fuzzing techniques for test generation. In this study, we  refrained from using LSCD to guide fuzzing, as employing the same criteria for both test generation and evaluation would influence the results.
We will investigate using LSCD to detect biases in DNN and drive retraining strategies that mitigate such problems. Finally, we aim to investigate the capability of LSCD to characterize DNN boundaries in the latent space of DNNs.

\bibliographystyle{spmpsci}
\bibliography{main}

\end{document}